\documentstyle[12pt]{article}
\newcommand{\be}{\begin{equation}}
\newcommand{\bea}{\begin{eqnarray}}
\newcommand{\eea}{\end{eqnarray}}
\newcommand{\ba}{\begin{array}}
\newcommand{\ea}{\end{array}}
\newcommand{\ee}{\end{equation}}

\expandafter\ifx\csname mathbbm\endcsname\relax

\else

\fi
\def\tQ1i{{\tilde Q}_i^1}
\def\tQ2i{{\tilde Q}_i^2}
\def\tQi{{\tilde Q}_i}
\def\tQ{{\tilde Q}}
\begin{document}
\begin{titlepage}
\hfill
\vbox{
    \halign{#\hfil         \cr
           hep-th/0204174 \cr
           IPM/P-2002/012 \cr
           SU-ITP-02/13 \cr
           } 
      }  
\vspace*{20mm}
\begin{center}
{\Large {\bf 
Strings in PP-Waves and Worldsheet Deconstruction  
}\\ }

\vspace*{15mm}
\vspace*{1mm}
{Mohsen Alishahiha$^a$ and Mohammad M. Sheikh-Jabbari$^b$}

\vspace*{1cm}

{\it $^a\ $ Institute for Studies in Theoretical Physics and Mathematics 
(IPM)\\
P.O. Box 19395-5531, Tehran, Iran}\\ 
{\tt {e-mail: alishah@theory.ipm.ac.ir}}\\
\vspace*{2mm}
{\it $^b\ $Department of physics, Stanford University\\
382 via Pueblo Mall, Stanford CA 94305-4060, USA}\\
{\tt {e-mail: jabbari@itp.stanford.edu}}\\

\vspace*{1cm}
\end{center}

\begin{abstract}
Based on the observation that $AdS_5\times S^5/Z_k$ orbifolds have a maximal 
supersymmetric PP-wave limit, the description of strings in PP-waves in 
terms of ${\cal N}=2$ quiver gauge theories is presented. We consider two 
different,  small and large $k$, cases and show that the operators in the 
gauge theory which correspond to stringy excitations are labelled by two 
integers, the excitation and winding or momentum numbers.
For the large $k$ case, the relation between the space-time 
and worldsheet deconstructions is discussed. We also comment on the 
possible duality between these two cases.

\end{abstract}

\end{titlepage}

\section{Introduction}

The PP-waves has been shown to be maximal supersymmetric solutions of type 
IIB supergravity \cite{Guven} and arise as  the Penrose limit
\cite{Penrose} of another maximal SUSY supergravity solution, namely 
$AdS_5\times S^5$ \cite{{KG},{Blau}}.\footnote{In the past two months 
there 
have been a cascade of papers devoted to this subject 
\cite{{List}}.} The (type IIB Green-Schwarz) string theory in the
PP-wave background has been formulated \cite{Met} and shown  to be exactly 
solvable \cite{MT}. The string $\sigma$-model action written in the 
light-cone takes a very simple form
\be\label{StringPP}
S={1\over 2\pi\alpha'}\int dt\int_0^{2\pi\alpha' p^+} d\sigma\ 
\left[{1\over 2} \dot{X}_i^2 -{1\over 2} {X'}_i^2- 
{1\over 2} \mu^2{X}_i^2+i\bar{S}(\Gamma^i\partial_i+\mu I)S\right]\ , 
\ee
where $X_i$ 's are eight transverse bosonic string worldsheet fields,  
$I=\Gamma^{1234}$ and $S$ is a Majorana (space-time) spinor. The parameter 
$\mu$, which is the mass parameter of the strings, is the field strength 
of self-dual 5-form of the background PP-wave solution.
The maximally SUSY PP-wave background has a $SO(4)\times SO(4)$ rotational 
symmetry, which is manifest in the action (\ref{StringPP}).

Since PP-waves are related to the $AdS_5\times S^5$, Berenstien, 
Maldacena and Nastase (BMN), have shown how  to translate the 
Penrose limit over the gravity side into the dual ${\cal N}=4$ SYM gauge 
theory \cite{BMN}. According to BMN dictionary, taking the Penrose limit 
corresponds to considering the sector of the gauge theory operators with 
large R-charge $J$ and large conformal weight $\Delta$, with $\Delta\sim 
J$ and $\Delta-J=positive\ and\ finite$. In fact BMN explicitly constructed 
the string 
creation-annihilation operators through particular gauge theory operators.
In this way they conjectured a way to obtain ``string bit'' formulation of 
the strings on PP-waves in the gauge theory language (the string bits 
correspond to certain chiral primary operators in the gauge theory).

Soon after the BMN work it was shown that the other supergravity solutions 
which are of the form $AdS_5\times {\cal M}^5$, where ${\cal M}$ is a 
smooth Einstein manifold and preserves some SUSY may also lead to the 
maximal supersymmetric PP-waves in the Penrose limit 
\cite{{GO},{IKM},{ZS}}. Since 
string theory on these backgrounds is dual to a ${\cal N}=1, D=4$ 
super-conformal field theory (SCFT), in the same spirit of the BMN work, 
there should be particular sub-sectors of the ${\cal N}=1$ gauge theories 
which shows the enhancement of SUSY and hence R-symmetry.

The PP-wave limit of the $AdS_5\times S^5/Z_k$ orbifold has also been 
considered \cite{{AS},{Orb1},{Orb2}}. In \cite{AS} we showed that
the $AdS_5\times S^5/Z_k$ orbifold, depending on how we take the Penrose 
limit, admits two PP-waves. One is  half supersymmetric and is 
the orbifold version of the maximal SUSY PP-wave background. However, the 
other one is the maximal SUSY PP-wave, as if there were no orbifolding.
String theory on $AdS_5\times S^5/Z_k$ is dual to a 
${\cal N}=2,\ D=4$,  $SU(N)^k$ quiver gauge theory with 
bi-fundamental matter fields.
In \cite{{AS},{Orb1}} the half SUSY case were considered and the 
stringy creation-annihilation of the string on the orbifolds of PP-waves
were constructed out of the proper dual (``holographic'') gauge theory 
operators. 

In this work we would like to consider the maximal SUSY PP-wave limit of 
the orbifolds and specify the operators in the dual ${\cal N}=2$ gauge 
theory side which correspond to strings on PP-waves. Therefore, 
strings on PP-waves would have descriptions in terms of ${\cal N}=4$
\cite{BMN}, ${\cal N}=1$ \cite{{GO},{IKM},{ZS}} and also ${\cal N}=2$
super-conformal gauge theories.

In the maximal SUSY PP-wave limit of the $AdS_5\times S^5/Z_k$ 
orbifolds $k$ remains a free parameter, i.e. the final supergravity 
background and hence the string theory do not depend on $k$ at all.
However, from the field theory side, $k$ is still there. 
In fact $k$ always appears as  $kJ$, where $J$ is the R-charge. 
In particular one may study small or large $k$ limits, while 
$kJ$ is fixed and large. The large $k$ limit of the orbifolds have been 
previously considered in \cite{{Nima1},{Nima2}} where the interesting 
effect of  ``deconstruction'' of space-time dimensions was introduced and 
studied. 

This article is organized as follows. In section 2 we briefly review the
results of \cite{AS} and show how the maximal SUSY PP-wave background 
arises in the orbifold case. In section 3, we consider the ${\cal N }=2$
quiver gauge theory  case and construct the stringy operators corresponding 
to strings
moving in PP-wave background. We consider two different cases of small 
and large $k$. We also show that in this quiver gauge theory language 
there is a room for another quantum number. This quantum number can be 
thought of as momentum (or winding) number for small (or large) $k$ cases. 
The  last section is devoted to discussion and conclusions.

\section{Review of the PP-wave limit of $AdS_5\times S^5/Z_{k}$}

In this section we briefly review the results of \cite{AS} where
it was shown that $AdS$ orbifolds admit the maximal SUSY PP-waves as the 
Penrose limit. To show how it works, consider the supergravity solution 
corresponding to $AdS_5\times S^5/Z_k$ solution   
\be
l_s^{-2}ds^2_{10}= R^2 \left 
(-\cosh^2\rho dt^2+d\rho^2+\sinh^2\rho 
d\Omega_3^2+ d{\cal M}_5^2 \right ) \ ,
\label{metric}
\ee
with
\be\label{calM}
d{\cal M}^2=d\alpha^2+\cos^2\alpha d\theta^2+{\sin^2\alpha \over 4}
(d\gamma^2+\sin^2\gamma\;d\delta^2)
+{\sin^2\alpha\over k^2}[
d\chi+{k\over 2}(\cos\gamma-1)d\delta]^2\ ,
\ee
where $\chi={\hat y}/l_s$ and $R^4=4\pi g_s N k$.
$\chi$ ranges from 0 to $2\pi k$ and the $Z_k$ orbifold is defined
by the identification $\chi\equiv \chi+2\pi$. The above metric can be 
understood as metric for $N$ D3-branes sitting at 
the orbifold $Z_k$, where  D3-branes are along $0123$ and 
$Z_k$ is acting on $4567$ directions (as usual) and $89$ is the fixed plane  
(the intersection of $89$ plane with $S^5$ is a circle parameterized by angle 
$\theta$). At the conformal point D3-branes are sitting at the 
orbifold fixed point ($\alpha=0$). 
The IIB self-dual five form is found to be
\be
F_{\chi\alpha\theta\gamma\delta}= 
(*F)_{\chi\alpha\theta\gamma\delta}= {R^4\over 
k}\alpha'^2\cos\alpha\sin^3\alpha\sin\gamma \ .
\ee
The (type IIB) string theory on the above background is dual to
a four dimensional ${\cal N}=2$ SYM theory with gauge group
$SU(N)^{k}$ and $k$ bi-fundamental matter
hypermultiplets {\it i.e.} $(Q_i, {\tilde Q}_i)$, $i=1,2,\cdots , k$,  in 
${\cal N}=1$ notation. The $Z_{k}$ acts as a permutation of the gauge 
factors.
The dimensionless gauge coupling of each $SU(N)$ part of the 
gauge group is $g_{YM}^2 \sim g_s k$.
Besides the scalars in bi-fundamental hypermultiplets we also have
the complex scalar in the vectormultiplet, $\varphi_i$.

The maximal SUSY PP-wave background can be obtained by focusing on
the particles moving along the $\chi$ direction. Namely, let us consider 
the scaling
\bea\label{limit}
t=x^+ +R^{-2}x^-, \ \ \ \ \ \ \   {1\over k}\chi=x^+ -R^{-2}x^-, \cr  
\rho={r\over R}\ ,\ \ \gamma={2x\over R}\ ,\ \ \ \alpha={\pi\over 
2}-{y\over R}\ ,
\ R\to\infty 
\eea
while keeping $x^+$, $x^-$, $r,\ x$ and $y$ fixed.
In this limit the metric (\ref{metric}) becomes \cite{AS}
\be
l_s^{-2}ds^2=-4dx^{+}dx^{-}-\mu^2{\vec z}^2dx^{+2}+d{\vec z}^2\ ,
\label{MAX}
\ee
and
\be
F_{+1234}=F_{+4567}=\mu\ . 
\ee
We would like to comment that, although the above solution looks exactly 
the same as the PP-wave considered by BMN, there is a difference, i.e. the 
$x^-$ direction in this case is compact. As we will show this will lead
to the possiblity of having another (integer) quantum numbers for the 
string states, namely the light-cone momentum and winding numbers.
 
This background, and consequently the strings moving in this background, 
do not depend on $k$  and in particular the above limit is true for 
both small and large $k$. 
As we see, the SUSY has been enhanced to the maximal 32 supercharges, and 
also the $SU(2)\times U(1)$ R-symmetry has been enhanced to $SO(4)\cong 
SU(2)\times SU(2)$. However, from the gauge theory side $k$ is still 
meaningful. Therefore in the proper gauge theory operators which describe 
the stringy operators, for the small $k$ case, $k$ can only appear as 
another quantum number and its corresponding operator should commute with all 
the string creation operators. We would like to note that in the PP-wave 
limit (\ref{limit}) we are focusing on the geodesics far from the orbifold 
point $(\alpha=0)$. {}From the gauge theory side this means that we are 
studying the theory far from the conformal point. This in particular 
implies that in the gauge theory operators corresponding to the string 
excitations only one of the $SU(N)$ gauge factors should be manifest.

\section{${\cal N}=2$ Description of strings in PP-waves}

To construct the string operators in the gauge theory language, let us 
first very shortly review the BMN arguments \cite{BMN}. The maximal SUSY 
type IIB PP-wave background has a $SO(4)\times SO(4)$ isometry, as well as 
translations along $x^+, x^-$ (and also sixteen other spatial and 
non-compact isometries \cite{Comp}).
These 
isometries from the gauge theory point of view correspond to the 
reminiscent of $SO(4,2)$ conformal symmetry and $SO(6)\cong SU(4)$ 
R-symmetry.
Choosing the sub-sector of the operators of the theory with large 
R-charge $J$ under a $U(1)$ factor of $SO(6)$, will effectively reduce 
the $SO(6)$ to $SO(4)$. On the other hand focusing on operators which are 
``almost'' chiral primary (i.e. operators of large conformal weight, 
$\Delta$ where $\Delta-J=finite$) we have also effectively reduced the 
conformal group to the $SO(4)$ subgroup \cite{Sumit}.
Although in the PP-wave  limit the 't Hooft coupling is very large, 
working with ``almost'' chiral primary operators has the advantage of many 
supersymmetric cancellations so that the effective gauge theory coupling 
is now 
\be\label{geff}
g^2_{eff}={g^2_{YM}N\over J^2}\simeq {1\over 4\pi}\ .
\ee

According to BMN the stringy operators can be written in terms of definite 
gauge theory operators. Let $Z$ be the (complex) scalar field which 
carries one unit of R-charge. The strings/gauge theory duality can be   
summarized as follows:

\begin{table}[htbp]
\begin{center}
        \begin {tabular}{|c||c|}
\hline
  Gauge Theory &  String Theory \\
\hline
  $Tr(Z^J)$ & vacuum  \\
     & \\
  $g_{eff}$ & ${1\over \alpha' \mu p^+}$  \\
         &  \\
$\sum_{l=1}^J\ e^{{2\pi i l n\over J}}\ Tr(Z^l \phi_a Z^{J-l})$ & 
$a^{\dagger\ a}_n$  \\
 & \\

$\sum_{l=1}^J\ e^{{2\pi i l n\over J}}\ Tr(Z^l D_aZ Z^{J-l})$ & 
$a^{\dagger\ (4+a)}_n$  \\
& \\
\hline
\end{tabular}
\caption{${\cal N}=4$ SYM-String Theory Correspondence}
\end{center}
\end{table}
\noindent
where $a=1,2,3,4$, $\phi_a$ are the four scalars (other than $Z$, 
${\bar 
Z}$), $D_a$ is the covariant derivative of the ``holographic'' 
gauge theory \cite{Holographic} and  $p^+$ is the strings light-cone momentum.

Now let us concentrate on the orbifold model, the case with 
half supersymmetry. Starting with a ${\cal N}=4$ SYM theory, the 
orbifolding will break $SU(4)$ R-symmetry to $SU(2)\times U(1)$, and 
the $Z_k$ is the discrete subgroup of the broken $SU(2)$ piece.  
{}From the string theory point of view, the $Z_k$ action is a symmetry of 
the action (\ref{StringPP}) and one can orbifold four of the 
$X_i$ components. Similar to the flat case, this orbifolding preserves 
half of SUSY, however, breaks the $SO(4)\times SO(4)$ symmetry to 
$SO(4)\times SU(2)$. Hence, to construct the corresponding operators from  
the gauge theory point of view, one should start with operators which 
carry the R-charge under the $U(1)_R$ part of the $SU(2)\times U(1)$ 
R-symmetry. More explicitly the sequence of $Z$'s should be replaced with
a proper combination of $\varphi_i$'s, the scalar in the vectormultiplets 
\cite{{AS},{Orb1}}. {}From the supergravity side one can understand this 
noting that the orbifold of PP-waves can be obtained from the 
$AdS_5\times S^5/Z_k$ solutions and focusing on the geometry 
seen by particles which are moving very fast near the orbifold point (or 
conformal point from the corresponding quiver gauge theory side) along the 
$S^1$ (in the $S^5$) which is fixed under orbifold action \cite{AS}.   

However, the situation for the maximal PP-wave limit of the $AdS_5\times 
S^5/Z_k$ is completely different.  As it is seen from (\ref{MAX}), 
the particle moving (spinning) very fast  along $\chi$ 
direction while sitting at $\rho=0,\alpha={\pi\over 2}$,  
feels a smooth metric. Since this metric is exactly
what we have in the non-orbifold case (of course except for the 
compactness of $x^-$),  one leads to the
conclusion that the $Z_k$ subgroup is replaced by a $U(1)$ subgroup and in 
addition, the two other generators of the broken $SU(2)$ combined with the 
$U(1)_R$ generates the new $SU(2)$ so that the R-symmetry is again 
$SO(4)\cong SU(2)\times SU(2)$. Furthermore, being far from the 
conformal point, instead of $SU(N)^k$ we effectively deal with a single 
$SU(N)$ gauge group. This is again what we expect to see from the BMN 
picture. Noting the scaling (\ref{limit})
the light cone momenta are
\bea\label{p+-}
2p^{-}&=&i\partial_{x^+}=i(\partial_t+k\partial_{\chi})=\Delta-kJ
\\ &&\cr
2p^{+}&=&i\frac{\partial_{x^-}}{R^2}=i
\frac{(\partial_t-k\partial_{\chi})}{R^2}=\frac{\Delta+kJ}{R^2}\ ,
\eea
where $J$ is the eigen-value for $-i\partial_{\chi}$ operator.
Observe that what is important and should be large is
the effective angular momentum $J_{\rm eff}=kJ$. In fact it is
this quantity which has to be taken large in the large $R$ limit namely, 
\be
(kJ)^2\sim R^4=4\pi g_s N k\ .
\ee

Note that the coupling of the effective $SU(N)$ gauge theory is of order of 
$g_s k$.  Hence the effective gauge coupling in our 
case (both large and small k cases) is
\be
g^2_{eff}={g^2_{YM}N\over (Jk)^2}\sim {g_s N k\over R^4}\simeq {1\over 
4\pi}\ .
\ee
That is, we can still treat the effective $SU(N)$ gauge theory 
perturbatively. Although in the string theory side $J$ and 
$k$ always appear as $kJ$, from the gauge theory point of view $J$ 
and $k$ still have their usual physical meanings. Here we study two 
different cases (in  the gauge theory side) which should correspond to 
the same string theory, namely, small $k$ (and large $J$) and large $k$ 
(small $J$) limits. 

\subsection{Small $k$ case}

As we have discussed, the $J_{eff}=kJ$ should be large and scale as $R^2$.
First we consider the case where $k$ is of order of one and $J\sim R^2$. 
In order to construct the chiral operators corresponding to the closed 
string excitation we first need to identify the vacuum state.

Let us define $Z_i=Q_i^1+iQ_i^2$. It is evident that $Z_i$'s are in the 
bi-fundamental representation, i.e. under gauge transformations 
$Z_i\to U_iZ_i U_{i+1}^{-1}$. The conformal dimension of $Z_i$ fields is 
one and their $\chi$ R-charge is ${1\over  k}$ (and hence their 
$J_{eff}=k\cdot{1\over k}=1$).  
Next we  define a scalar field $Z$, 
\be\label{Z}
Z=\prod_{i=1}^{k} Z_i
\ee
which has dimension $k$ and  therefore
\be\label{vac}
{\cal O}_{vac}={\rm Tr}(Z^J)
\ee
is chiral primary, i.e. ${\cal O}_{vac}$ has $\Delta-kJ=0$.  
We would like to note that the $Tr$ in the above expression is made over
$N\times N$ matrices. These matrices are in fact in the adjoint
representations of only one of the $SU(N)$ factors of our $SU(N)^k$ gauge
theory. This $SU(N)$ factor have been chosen to be the one parameterized by
$U_1$. It is easy to check that ${\cal O}_{vac}$ do not depend on this
specific choice of the $SU(N)$ factor. 
To see that, it is convenient to
define the ``twist'' operator, $\omega$
\be\label{twist}
\omega F_i \omega^{-1}= F_{i+1}
\ee
where $F_i$ is a generic operator of the $i$'th factor of the $SU(N)^k$
gauge group.  If the above state corresponds to the string theory vacuum 
state (or more precisely the state with zero excitation number), 
we see that it is $k$-fold degenerate. 

Similar to the BMN case, the idea is to discretize the string worldsheet 
into $J$ pieces. However, noting the operator corresponding stringy vacuum 
state, at each point of the string bit we have a $k$-point ($Z_k$ lattice) 
internal space. Then to obtain the string oscillatory modes we need to 
insert proper operators into the $J$-point string of $Z$'s.

Let us define the operators
\bea\label{phia}
{\phi}^{1}= \prod_{i=1}^k{({\varphi}_i^1)}Z_i\ &,\ &
{\phi}^{2}= \prod_{i=1}^k{({\varphi}_i^2)}Z_i\ \ ,\ \\
{\phi}^{3}=\prod_{i=1}^k{\tilde Q}_i^1\ & ,\ &
{\phi}^{4}=\prod_{i=1}^k{\tilde Q}_i^2\ ,
\eea
where ${\varphi}_i^{1,2}$ are the two real scalars of the  complex adjoint
scalar of the vectormultiplet. We note that each of these scalars
$\phi^a$ ($a=1,2,3,4$)
has $\Delta-kJ=k$ and is in the adjoint representation of the first $SU(N)$ 
gauge factor. Hence, they are the proper operators needed for constructing 
stringy excitations.
It would be more convenient to use $kN\times kN$ notation. In this notation
the operators we are dealing with are of the form
\be
{\varphi}=
\left(\matrix{
\varphi_1 &     &       &       &      &  \cr
    & \varphi_2 &       &       &      &  \cr
    &     &  \varphi_3  &       &      &  \cr 
    &     &       &\quad\cdot&      &     \cr
    &     &       &       &\quad\cdot& \cr
    &     &       &       &       & \varphi_k
}\right)\ \ ,
{\cal{D}}_a=\left(\matrix{
{D_a}_1 &     &       &       &      & \cr
    & {D_a}_2 &       &       &      & \cr
    &     &  {D_a}_3  &       &      & \cr
    &     &       &\quad\cdot&      & \cr
    &     &       &       &\quad\cdot& \cr
    &     &       &       &       & {D_a}_k
}\right)\ \ ,
\ee
all the off-diagonals are zero, and each element is a $N\times N$
matrix. In the above ${D_a}_i$ is the covariant
derivative corresponding to the $i$'th $SU(N)$ factor. And the 
bi-fundamentals in the hypermultiplet as $kN\times kN$ matrices are
\be
{\cal Z}=
\left(\matrix{
0           & Z_{1} &             &       &       &        \cr
            & 0           & Z_{2} &       &       &        \cr
            &             & 0           & \quad\cdot &       &        \cr
            &             &             & \quad\cdot & \quad \cdot &    \cr
            &             &             &       & 0 & Z_{k-1}  \cr
Z_{k} &             &             &       &     &  0
}\right),\ 
\tilde{{\cal Q}}=
\left(\matrix{
0           &    &             &       &       &   {\tilde Q}_k   \cr
{\tilde Q}_1             & 0           &  &       &       &        \cr
            & {\tilde Q}_2     & 0           &    &       &        \cr
            &             &  \quad\cdot & \quad \cdot &     &   \cr
            &             &        &\quad\cdot & 0   &  \cr
              &             &       &       &{\tilde Q}_{k-1}& 0}\right)\ .
\ee
Then instead of products over the $i$ indices one can use determinants over   
the $k\times k$ matrices, e.g. $Z=det{{\cal Z}}$. In this notation 
\bea\label{phis}
{\phi}^{1}= det({\varphi}^1{\cal Z}) &,\ &
{\phi}^{2}= det({\varphi}^2{\cal Z}) \cr
{\phi}^{3}=det({\tilde {\cal Q}}^1)\ & ,\ &
{\phi}^{4}=det({\tilde {\cal Q}}^2)\
\eea
Also we define
\be
D_a Z\equiv det({\cal {D}}_a{\cal Z})\ .
\ee
These scalars together with the covariant derivative of $Z$
will give us the basis for constructing closed string excitation.
The corresponding operators with
$\Delta-kJ=k$ are given by
\bea\label{JJJ}
{a^{\dagger}_{(q)n}}^a\longleftrightarrow 
{\cal O}_{(q)n}^a&=&\sum_{p=1}^k\sum_{m=1}^J \omega^p\left[{\rm 
Tr}(Z^m
\phi^a Z^{J-m})\right] \omega^{-p}\ 
e^{{2\pi i mn\over J}}\ e^{{2\pi i pq\over k}}\
 \cr
&&\cr
{a^{\dagger}_{(q)n}}^{4+a}\longleftrightarrow 
{\cal O}_{(q)n}^{4+a}&=&\sum_{p=1}^k\sum_{m=1}^J \omega^p\left[{\rm Tr}(Z^m
{D_a Z} Z^{J-m})\right]\omega^{-p}\ 
e^{{2\pi i mn\over J}}\ e^{{2\pi i pq\over k}}\ ,
\eea
with $a=1,2,3,4$. As we see the above operators have two type of indices, 
$n$ and $q$. The $q$ indices show the ``momentum'' quantum number
with respect to $\omega$, i.e.
$$  
\omega {\cal O}_{(q)n}^a \omega^{-1}= e^{{-2\pi i q\over k}} 
{\cal O}_{(q)n}^a\ .
$$
Note that operators ${\cal O}_{(q)n}^a$ commute with $\omega$, for any $n$ 
and $q$. Therefore index $q$ is an internal quantum number.

The above operators are zero due to the cyclicity of the trace.
Similar to the BMN case, these operators can be used to construct the
excitations of the the closed strings in the 8 transverse directions.
This can be done by a further $\phi^a$ or $D_a Z$ insertion. For example
\be
{a^{\dagger}_{(q)n}}^a\ {a^{\dagger}_{(q)-n}}^b \longleftrightarrow 
{\cal O}_{(q)n}^{ab}\sum_{p=1}^k\sum_{m=1}^J 
\omega^p\left[{\rm 
Tr}(\phi^a Z^m \phi^b Z^{J-m})\right] \omega^{-p}\ 
e^{{2\pi i mn\over J}}\ e^{{2\pi i pq\over k}}\ .
\ee
It is remarkable that 
\be
\omega {\cal O}_{(q)n}^{ab} \omega^{-1} = e^{{-2\pi i q\over k}} 
{\cal O}_{(q)n}^{ab}\ . 
\ee
Therefore we can construct states whose $q$ momentum are non-zero, however, 
the sum of the other quantum number, $n$, for any closed string state should 
be zero, similar to the BMN case \cite{BMN}. One can also write down the 
fermionic 
operators using the supersymmetry present in the theory.

As we expect, these operators have the same structure as in the ${\cal
N}=4$ case. However, one should note that $\Delta-kJ$ for these operators is
$k$ (instead of one). This can be understood as follows:

In the Penrose limit which leads to the maximal SUSY case
we are considering geometry seen by particle far from the orbifold
fixed point ($\alpha=\pi/2$). In other words from that particle point of
view it seems that we are working in the covering space of the orbifold and
hence to compare our results with that of  BMN we should rescale
$p^+$ by a factor of $k$. This in particular implies that the dimension 
of the operators in the subsector of the theory we are focusing on, is 
always a multiple of $k$.  More intuitively, in the BMN picture
the length $R$ have been divided into $J$ pieces and since $J\sim R^2$ the
momentum along this direction should be measured in ${1\over R}\sim {R\over
J}$ units. However, in our case ${kJ\sim R^2}$, while still we have divided
the length $R$ into $J$ pieces. Therefore the momentum units are now
${k\over R}$\ .
To summarize, taking a fast moving particle in the $\chi$ 
direction breaks the Lorentz group $SO(2,4)$ to $SO(4)$ and effectively 
reduces the gauge group $SU(N)^k$ to $SU(N)$ while enhances the R-symmetry 
from $U(1)_R\times SU(2)_R$ to a global $SO(4)$ symmetry.

\subsection{Large $k$ case}

In this part we focus on the large $k$ case. 
Let us first consider $J=1$ case while $k$ is scaling like $R^2$. 
Here the closed string worldsheet is (de)constructed from $k$-points which 
are $k$ equidistant points around a circle of radius $R$ (the distance 
between the string bits is going to zero as ${1\over R}$ in the large $R$, 
PP-wave, limit).  
The operator (\ref{vac}) (with $J=1$) corresponds to the vacuum of string 
theory. Then the stringy creation operators can be found by insertion of 
$\varphi_i$'s (scalars in the vectormultiplet), $\tQi$'s (other 
bi-fundamental scalars), and the covariant derivative in the sequence of 
$Z_i$'s in (\ref{vac}). More explicitly:
\bea\label{adagger}
{\cal O}^{1,2}_q &=& \sum_{p=1}^k e^{{2\pi i pq\over k}}  
STr(Z_1 \cdots Z_{p-1} \varphi_p Z_p \cdots Z_k) \cr
{\cal O}^{3,4}_q &=& \sum_{p=1}^k e^{{2\pi i pq \over k}}  
STr(Z_1 \cdots Z_{p-1} Z_p {\tilde{Q}}_p Z_p Z_{p+1} \cdots Z_k) \cr
{\cal O}^{4+a}_q &=& \sum_{p=1}^k e^{{2\pi i pq\over k}}  
STr(Z_1 \cdots Z_{p-1} D_a Z_p \cdots Z_k),  
\eea
where $a=1,\cdots 4$. In the above $STr$ has been defined as
\bea\label{STr}
STr(A_1A_2\cdots A_{p-1}B_p\cdots A_k)&=&
{1\over k}\sum_{p=1}^k Tr(A_1A_2\cdots A_{p-1}B_p\cdots A_k) \cr
&=& {1\over k}\sum_{p=1}^k \omega^p Tr(B_1A_1A_2\cdots A_k) \omega^{-p} \ , 
\eea
where $\omega$ is the twist operator. The $STr$ defined as above, guarantees 
the necessary ``cyclicity'' condition of the trace, crucial for 
constructing the stringy  operators. Therefore the operators (\ref{adagger}) 
are identically zero. 
Using $STr$ is further justified noting that $STr$ is actually the trace 
over the diagonal $SU(N)$ factor and the fact that 
the maximal supersymmetric PP-wave limit corresponds to
quiver gauge theory far from its conformal point. Identifying 
(\ref{vac}) with the string theory vacuum state gives a non-zero 
expectation value to the vacuum operator. This in particular implies 
that the original $SU(N)^k$ gauge theory is higgsed down to the diagonal
$SU(N)$ and hence the traces should be made over that subsector only.
One can proceed to construct closed 
string operators by insertion of another  $\varphi_i$ or $\tQi$ in the 
$Z_i$ chain.
\vskip .5cm
{\bf {\it Relation to space-time (de)construction}}
\vskip .5cm
An interesting feature of our model in the large $k$ and worldsheet 
deconstruction, is its close 
relation with space-time
deconstructing dimension scenario \cite{Nima1}. In fact the model we are
considering here is exactly the one studied in \cite{Nima2} which exhibits 
a 
new dimension for large $k$ limit. In \cite{Nima2} it has been shown that 
the 
quiver model we have been studying here in its Higgs branch flows to a five 
dimensional $SU(N)$ gauge theory. In our
point of view it can be understood through T-duality where our solution 
will be
a configuration of NS5 and wrapped D4-branes. The
supergravity solution of this model, as well as its uplift to M-theory, 
have
been considered in \cite{{ALIOZ}}. The theory has a Higgs phase where
the gauge group $SU(N)^k$ breaks to its diagonal $SU(N)$ factor and hence  
is effectively the theory on the worldvolume of D4-branes wrapped on a
circle with the radius proportional to $k$.  

To establish the relation between worldsheet and space-time deconstructions
we note the usual KK expansion of a generic field of five dimensional 
theory reduced on a circle, i.e.
$$
A({\vec X},y)=\sum_{p=1}^{\infty} A_p({\vec X}) e^{ipy}\ ,
$$
where ${\vec X}$ are parametrizing the four dimensional part and 
$y$ which is ranging from 0 to $2\pi$, the compact fifth direction. 
Now if we replace the circle with a discretized version consisting of $k$
points, $p$ in the KK expansion ranges from 1 to $k$ and $y$ can be 
replaced with ${2\pi q\over k}$,  $q$ ranging from 1 to $k$. In this 
discretized 
form  the usual KK expansion is essentially the same as the expansion
we have used to construct the stringy operators (cf. Eq.(\ref{adagger})).   
This could lead to a close relation between string theory on the PP-wave and 
this five dimensional theory. Definitely this consideration deserves more 
study and we are going to push this direction in our future work 
\cite{Progress}.

\vskip .5cm
{\bf {\it $J>1$ case }}
\vskip .5 cm

Here we will show that $J>1$ opens the possibility for strings to have 
``winding''  modes. To visualize this, we recall the usual 
string windings. Let $\sigma$ be the closed string worldsheet parameter 
which ranges from $0$ to $2\pi$. Then, the winding number $w$ is the number 
of times a string repeats itself along a compact direction in the $2\pi R$ 
units ($R$ is the radius of the compact direction), i.e. $X(2\pi)-X(0)=2\pi 
R w$, where $X$ is the string worldsheet field. 
Usually the winding number, unlike the momentum modes, is not associated 
with a (quantum) operator and basically quantization of winding number is a 
classical effect. However, one can think of winding to be eigen-value of
an ``identity'' operator $\Omega$, where $\Omega X(0) \Omega^{-1}=X(2\pi)$.
Essentially $\Omega$ is translation operator on the worldsheet by $2\pi$ 
and $\Delta_{\Omega}X=2\pi R w$.
Now let us go back to the gauge theory case.   As we have shown 
for $J=1$ case we (de)construct worldsheet from $Z_i$ $1\leq i\leq k$. In 
other words index $i$ plays the role of worldsheet coordinate $\sigma$.
In particular twist operator $\omega$ generates infinitesimal translations 
along the worldsheet. 
However, to see the role of $J$ it is simpler to extend the range of $i$ 
index from $k$ to $kJ$ and then in the end we re-identify $Z_i$ with 
$Z_{i+k}$. To avoid possible confusions we will denote this extended index 
by $r$ $(1\leq r\leq kJ)$. Then any $r$ can be written as $r=mk+q$ where
$0\leq m\leq J-1$ and $1\leq q\leq k$. Similar to the previous case we use 
the operator $\omega$ to move in the $q$ part of the index $r$, moreover we  
introduce the new operator $\Omega$ which changes $r$ in $k$ units:
\bea\label{Oomega}
\Omega Z_r \Omega^{-1}&=&Z_{r+k},\ \ \ \Omega^{J}\equiv 1\ , \cr
\omega Z_r \omega^{-1}&=&Z_{r+1},\ \ \ \omega^{k}\equiv 1\ ,\ Z_{r}\equiv 
Z_{r+kJ}\ .
\eea 

Equipped with the above operators, $\omega$, $\Omega$ and the extended index 
$r$, we are ready to write the proper operators we are looking for. The 
proper vacuum state is still given by (\ref{vac}). This vacuum is $J$-fold 
degenerate (it is invariant under $\Omega$). The operators corresponding to 
excited strings can be defined as
\be\label{largek}
{\cal O}_{(n)q}^{\alpha}=\sum_{p=1}^k\ \sum_{m=0}^{J-1}
STr(Z^m\ Z_1^{\alpha} \ Z^{J-m-1}) 
e^{{2\pi ipq\over k}} e^{{2\pi imn\over J}}\ ,\ \ \ \alpha=1,2,\cdots, 8, 
\ee
where $Z$ is defined by (\ref{Z}) and
\be
Z^m\equiv \prod_{s=0}^{m-1} \Omega^s Z \Omega^{-s}\ ,
\ee
\be\label{XX}
Z_1^{\alpha}=\Omega^m (X_1^{\alpha}Z_1\cdots Z_k) \Omega^{-m}\ .
\ee
In the Eq.(\ref{XX}) $X_1^\alpha$ can be chosen among ${D_{a}}_1, 
\varphi_1^1,  \varphi_1^2, {\tilde Q}_1^1$ and  ${\tilde Q}_1^2$.
$STr$ is defined by (\ref{STr}). We should emphasize that after making all 
the manipulations again we set $Z_{r+k}\equiv Z_r$ (or equivalently,
$\Omega$ is set to identity).  

Noting the operator (\ref{largek}) following remarks are in order:
\vskip .5cm
{\it i)} As previous cases, the $STr$ is over the $N\times N$ matrices of 
the diagonal $SU(N)$ gauge factor. So, effectively the subsector of the 
operators we are dealing with has a single $SU(N)$ gauge symmetry. 
\vskip .5cm
{\it ii)} $\ \Omega {\cal O}_{(n)q}^{\alpha} \Omega^{-1}= e^{-{2\pi in\over 
J}} {\cal O}_{(n)q}^{\alpha},\ n=0,1, \cdots, J-1$.  
This in particular implies that $n$ is another quantum number related to 
strings center of mass degrees of freedom and can be identified with the 
winding number.
\vskip .5cm
{\it iii)} Because of the cyclicity of the trace, operator (\ref{largek}) is 
zero. The higher stringy excitations can be found by more insertions of 
$Z_1^\alpha$ 
(operators defined in Eq.(\ref{XX})).
\vskip .5cm
{\it iv)} The same extension for the range of indices from $k$ to $kJ$ can 
also be 
used  for the large $J$, small $k$ case. However, then the proper 
decomposition of index $r$ is $r=qJ+m$, $(0\leq q < k,\ 0\leq m < J)$.

\section{Discussions}

In this paper we have studied the Penrose limit of $AdS_5\times
S^5/Z_k$ which leads to the maximally supersymmetric PP-wave
background in type IIB string theory. Since the PP-wave we have found
from orbifold model has maximal SUSY,  we would expect to find
a subsector of ${\cal N}=2$ $SU(N)^k$ quiver theory which exhibits
the supersymmetry enhancement. In fact we have shown that there is such a 
subsector which is parameterized by conformal weight $\Delta$ and R-charge $kJ$ 
such that both $\Delta$ and $kJ$ are parameterically large while their 
difference $\Delta-kJ$ is finite.

The important point in this case is that the quantity which has to be taken 
large in the large $R$ limit is $kJ$ and not the angular momentum of the 
particle $J$. In fact, as far as the oscillator modes of the string theory 
(and also the background supergravity) are concerned 
$J$ and $k$ always appear as $kJ$. Therefore for large
$kJ$ limit we have to different choices: we can take either $k$ large and 
finite $J$ or large $J$ and finite $k$. 

For both cases we have studied the gauge invariant chiral operators in the 
quiver model and identified them with the string excitations of type IIB 
string theory on the maximally supersymmetric PP-wave. This subsector
of the quiver model has the same structure as the one considered
for ${\cal N}=4$ \cite{BMN} indicating that there is a supersymmetry 
enhancement and furthermore the $SU(N)^k$ gauge symmetry is higgsed down to 
the diagonal $SU(N)$.

We note, however, that although the PP-wave solution looks very much like 
the one considered in \cite{BMN}, there are some differences which have to 
be taken into account and these differences make the quiver 
model have richer structure. The first difference we would like to 
mention is: the direction which is combined with $t$ is compact and hence
the maximally supersymmetric PP-wave we find in this case has compact $x^-$ 
direction.
Moreover, since in the gauge theory side we have two quantum
numbers $k$ and $J$, the states in the string theory should have
an other extra quantum number in comparison to the ordinary
maximally PP-wave background studied by BMN. 
Therefore the operators we introduced here are labelled by two integers, the 
excitation and winding or momentum numbers. We have argued that in the large 
$J$, finite $k$ case there is an operator which commutes with all the 
stringy oscillators and its eigen-values (which range from $0$ to $k-1$) 
can be identified as momentum modes. For the large $k$, finite $J$ case, 
however, we have shown that the other (quantum) number, which ranges from 
$0$ to $J-1$, corresponds to the winding modes. 

We have also discussed the relation between the deconstruction of 
space-time dimensions through the ``theory space'' \cite{{Nima1},{Nima2}} 
and our worldsheet deconstruction. It would be quite interesting to make this 
arguments more explicit and also extend the same argument to the 
M-theory, similar to that of Ref. \cite{Nima2}. 
 
The other interesting open question we address here is the possible duality 
between the large $J$ and large $k$ cases. Besides the similarities in the 
form of the operators in these two cases (Eqs.(\ref{JJJ}) and 
(\ref{largek})), 
there is the usual T-duality which exchanges winding and momentum modes.
We also expect to see such a duality here, under which the $k$ and $J$
parameters are replaced while their product is kept fixed and scale
like $R^2$. We hope to come back to this question in future publications.

{\bf Note added}: While we were preparing our paper for submission we 
received the paper \cite{MV} where the same question has been studied. 

\vskip .5cm

{\bf Acknowledgments}

We would like to thank Farhad Ardalan, Hesam Arfaei, Nima Arkani-Hamed, 
Sunil Mukhi and Leopoldo Pando-Zayas and especially Keshav Dasgupta for 
fruitful discussions. 
The research of M. M. Sh-J. is supported in part by NSF grant PHY- 9870115 
and in part by funds from the Stanford Institute for Theoretical Physics.

\end{document}